**A. V. Yurkin**

A. M. Prokhorov General Physics Institute of RAS, e-mail: alvl1yurkin@rambler.ru

# RAY SYSTEM IN LASERS, NONLINEAR ARITHMETIC PYRAMID AND NONLINEAR ARITHMETIC TRIANGLES

The paper describes a system of rays declining at small angles in lasers. The correlation between a group of rays and binomial coefficients is shown. The correlation of distribution of rays in the system of numbers placed in a three-dimensional table, the nonlinear arithmetic pyramid is shown. Two types of nonlinear arithmetic triangles are considered. Various types of partitions of integers is described.

## 1. Introduction

In order to describe processes in lasers equipped with resonators, usually wave and geo-metro-optical (ray) models and equivalent light guide schemes are used [1, 2] .

The work [3] offers an illustrative geo-metro- optical model founded on the consideration of binomial distribution to describe the propagation of light in a laser provided with a multi-lobe mirror. Such mirrors were discussed in [4, 9] and consist of many semitransparent flat planes inclined towards the laser axis at a small angle $\gamma$ and turned symmetrically around the axis. Light from such a mirror is reflected conically as hollow cones, for practical reasons the mirrors themselves being considered to be thin. In the plane (two-dimensional) case the ray reflected from our mirror bifurcates. This paper considers the case. It is of interest to note that the notion of "ray" is equivalent to the notion of "plane wave" and light energy emanating from the laser is proportional to the number of rays.

In our work we use simple laws of geometric optics to describe in more detail our equivalent light-guide scheme, a branching system of rays inclined at small angles in lasers. This makes it possible to describe the processes taking place in lasers more thoroughly.

## 2. Bifurcation of rays and binomial distribution

In works [3, 5, 6] we studied branching systems of rays consisting of ray groups joined in polygonal paths consisting of links.

Let us consider the system of rays in more detail. The rays are inclined at small angles $\pm 2k\gamma$, where $k = 0,1,2,\dots$. All the group of rays may be formed from one ray in successive construction.

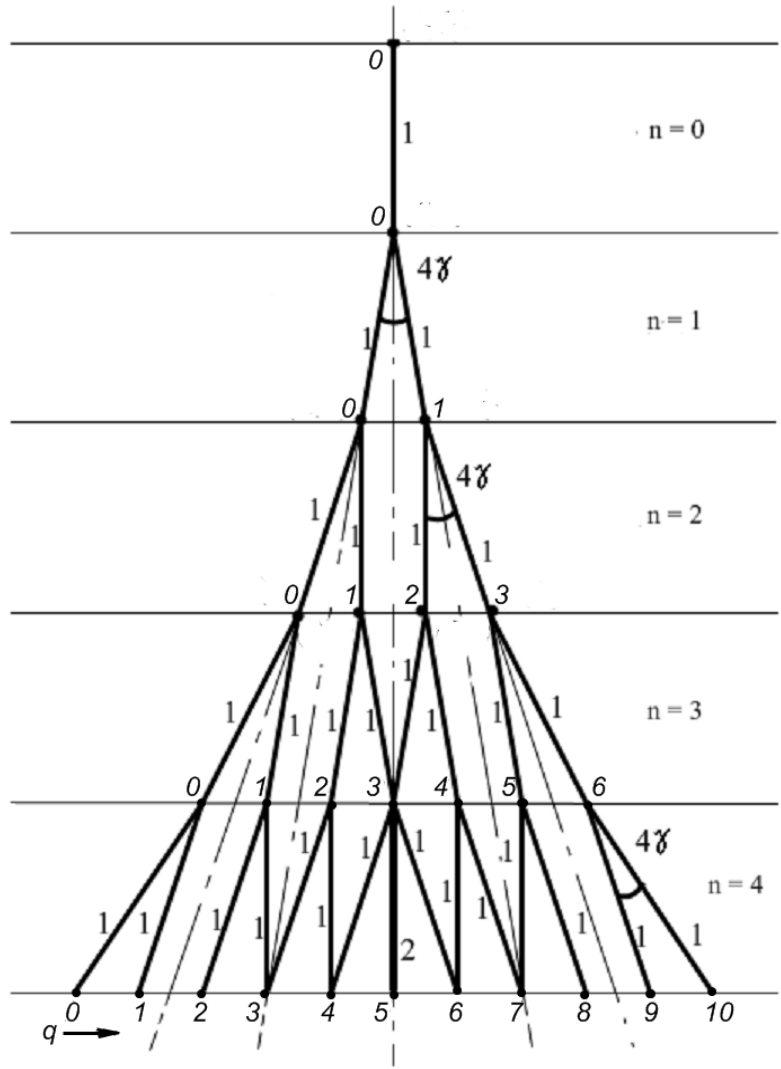

Fig. 1. The system of rays inclined at angles $\pm 2k\gamma$ with the vertical. The rays and links are shown by continuous lines, symmetry axes by line-and-dot lines. Light propagates from the top down. $n$ – the number of ray passage, $q$ – the number (italic type) of ray branching points.

Fig. 1 shows the part of the plane system [3] for ray group $K$, inclined at angles $\pm 2k\gamma$.

After each passage of light $n$, where $n = 0, 1, 2, 3\ldots$ (Fig. 1) rays are branched into two when reflected from the multi-lobe mirror. The initial passage (without branching) is considered to be null $(n = 0)$. The distribution of light energy outgoing along the rays is proportional to the number of rays. In the first three passages energy is distributed among the rays equally.

Some rays are parallel to each other, i.e. are inclined at equal angles.

$K$ rays propagate along $N$ links.

Beginning with the fourth passage $(n = 4)$, some rays are superimposed on each other, the distribution of energy among the rays becomes unequal. In the general case, the number of $K$ rays is greater or equal to the number of links $N$, i.e. $K \geq N$. Thus, for example. it is seen from Fig. 1 that $K = N = 2$ for $n = 1$; $K = N = 4$ for $n = 2$; $K = N = 8$ for $n = 3$; but $K = 16$, $N = 15$ for $n = 4$. In Fig. 1 the amount of $K$ rays is shown by figures and points $q$ of ray branching are shown by italic.

### 3. The Pascal triangle

It is known [7] that binomial coefficients $\begin{pmatrix} n \\ p \end{pmatrix}$ can be calculated by means of a two-dimensional numerical table, Pascal arithmetic triangle in which numbers are arranged by layers of rows where the binomial power $n$ is the ordinal number of the row beginning with the apex of the triangle and $p$ is the ordinal number of the number in the row and $0 \leq p \leq n$ (Fig. 2).

Fig. 2. Pascal triangle, two images: a) – symmetric, b) – non-symmetric; the sums of numbers in the rows are shown on the right.

Let us set the initial conditions for the number of the zero row $(n = 0)$:

$$\binom{0}{p} = 1, \tag{1}$$

for $(p = 0)$, i.e.

$$\binom{0}{0} = 1 \text{ and } \binom{0}{p} = 0 \tag{2}$$

for the others $p$.

Let us set also the boundary conditions for numbers $p$ of the other $n$-rows:

$$\binom{n}{p} = 0 \tag{3}$$

for $p < 0$, $p > n$.

Then the rule of successive filling Pascal triangle with numbers will be:

$$\binom{n}{p} = \binom{n-1}{p-1} + \binom{n-1}{p}, \text{ or } \binom{n+1}{p} = \binom{n}{p-1} + \binom{n}{p}. \tag{4}$$

The numbers in each of the lower rows in Pascal triangle, binomial co-efficients, are thee sum of two numbers of the upper row. The sum of coefficients for power $n$ of the binomial is $2^n$, shown in the column on the right in Fig. 2b.

### 4. Non-linear arithmetic pyramid

Three-dimensional generalizations of Pascal triangle are known as linear arithmetic pyramids, e.g. tetrahedron [7].

For our calculations let us also construct a tree-dimensional numerical table, i.e. non-linear arithmetic pyramid (Fig. 3) consisting of two-dimensional layers $n$ (Fig.4), where $n = 0,1,2,3\ldots$.

Let us find correspondence between the system of rays in Fig. 1 and the system of numbers located in the non-linear arithmetic pyramid in Fig. 3, 4.

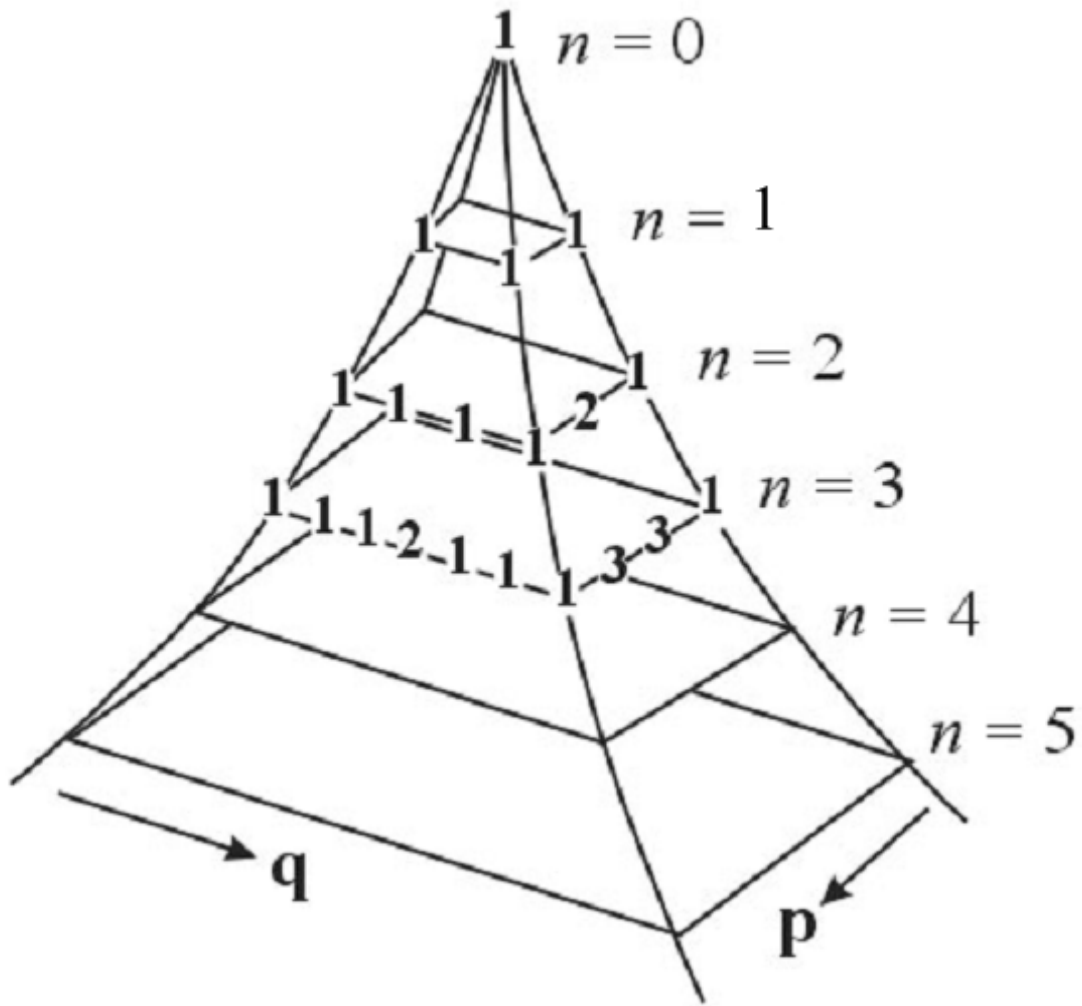

Fig. 3. Non-linear arithmetic pyramid. The number in the zero $(n=0)$ layer is unity, each following $n$-layer of the pyramid beginning with $n=1$ is filled with natural numbers and zeroes and has the form a rectangle with the height of $p+1$ and breadth of $q+1$.

The ordinal number of $n$-layer of the pyramid beginning with the apex corresponds to the passage of $n$ rays (Fig. 1), $p$ – row number, $q$ – the number of the column of the rectangular layer $n$ of the pyramid. With increasing $n$ number, $p$ grows linearly: $0 \le p \le n$, and $q$ – non-linearly: $0 \le q \le n(n+1)/2$ (Fig. 3, 4). number $p$ corresponds to the value of inclination of rays (Fig. 1) at angles multiple of $2\gamma$, number $q$ corresponds to the numbers of points of ray branching (Fig. 1).

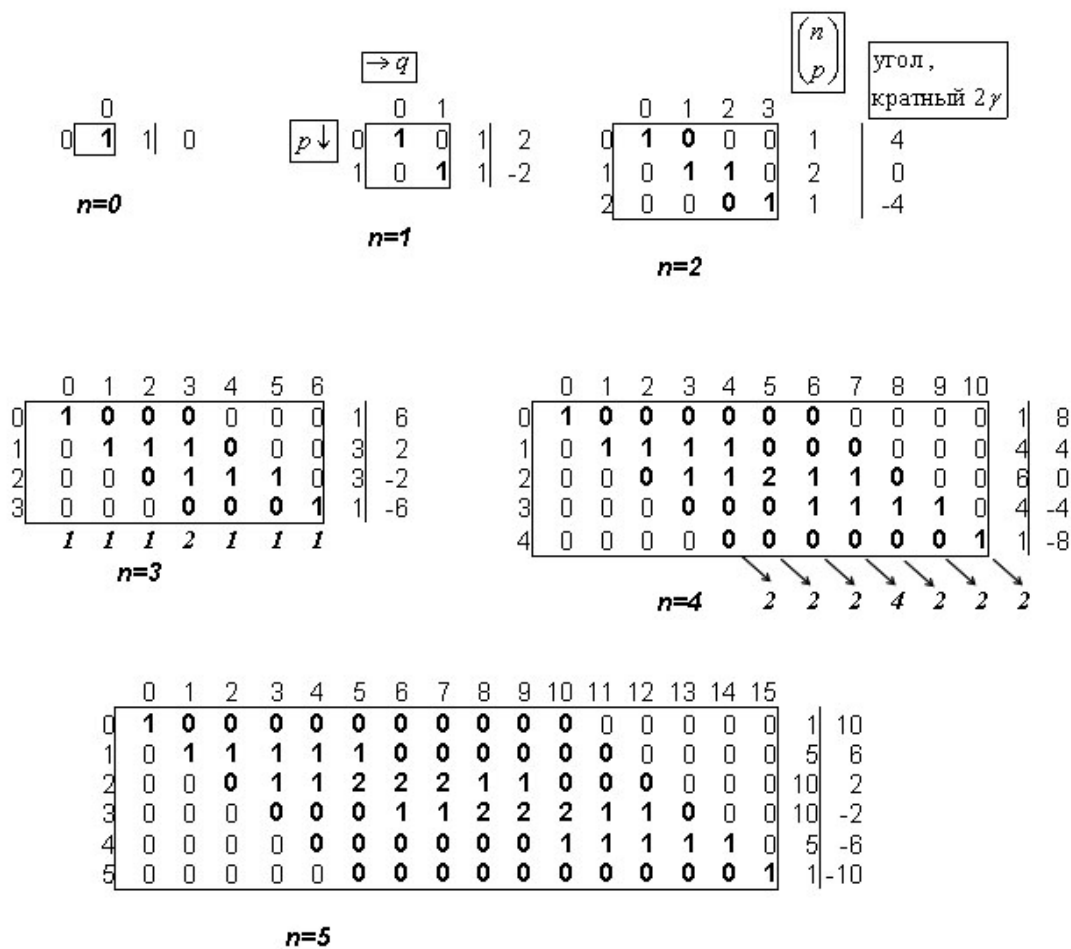

Fig. 4. The layers of the non-linear arithmetic pyramid for various values of $n$. $p$ – numbers of rows and $q$ – the numbers of columns for each of the layers. In the first column of figures beyond the rectangles on the right – the sums of row numbers $p$ equal to binomial coefficients; in the second column on the right values of ray inclination angles for each passage $n$ (Fig.1) are shown, which are multiple of angles $\pm 2\gamma$ or, to be more precise $2(n-2p)\gamma$. For $n = 3$ and $n = 4$ the sums of numbers are shown in vertical and inclined (45°) columns, respectively; these sums are shown in italics in lines below rectangles.

Let us denote the number located in $n$ – layer of the pyramid in row $p$ and column $q$ as

$$\begin{pmatrix} n \\ p \\ q \end{pmatrix} \qquad (5)$$

Let us set the initial value for the number of zero $(n = 0)$ layer as

$$\begin{pmatrix} 0 \\ p \\ q \end{pmatrix} = 1 \qquad (6)$$

for $p = 0, \; q = 0,$ . i.e

$$\begin{pmatrix} 0 \\ 0 \\ 0 \end{pmatrix} = 1 \;\; \text{and} \;\; \begin{pmatrix} 0 \\ p \\ q \end{pmatrix} = 0 \qquad (7)$$

for the other $p$ and $q$.

Let us also set the boundary conditions for the numbers $p$ and $q$ of the other $n$ - layers:

$$\begin{pmatrix} n \\ p \\ q \end{pmatrix} = 0 \qquad (8)$$

for $p < 0, \; p > n$ and $q < 0, \; q > n(n+1)/2$

Then the rule of successive filling with numbers of our three-dimensional table (Fig. 3, 4), beginning with the pyramid apex will be:

$$\begin{pmatrix} n \\ p \\ q \end{pmatrix} = \begin{pmatrix} n-1 \\ p-1 \\ q-p \end{pmatrix} + \begin{pmatrix} n-1 \\ p \\ q-p \end{pmatrix}. \qquad (9)$$

For example, for the first layer $(n = 1)$ for $p = 0, \; q = 0$:

$$\begin{pmatrix} 1 \\ 0 \\ 0 \end{pmatrix} = \begin{pmatrix} 0 \\ 0\text{-}1 \\ 0\text{-}0 \end{pmatrix} + \begin{pmatrix} 0 \\ 0 \\ 0\text{-}0 \end{pmatrix} = \begin{pmatrix} 0 \\ \text{-}1 \\ 0 \end{pmatrix} + \begin{pmatrix} 0 \\ 0 \\ 0 \end{pmatrix} = 1, \;\; \text{because} \;\; \begin{pmatrix} 0 \\ -1 \\ 0 \end{pmatrix} = 0, \;\; \text{and} \;\; \begin{pmatrix} 0 \\ 0 \\ 0 \end{pmatrix} = 1$$

from the condition (6), (7) and (8).

The other three numbers filling the first $(n = 1)$ pyramid layer are obtained in a similar way —

$$(p = 1, \ q = 0): \begin{pmatrix} 1 \\ 1 \\ 0 \end{pmatrix} = 0, \ (p = 0, \ q = 1): \begin{pmatrix} 1 \\ 0 \\ 1 \end{pmatrix} = 0, \ (p = 1, \ q = 1): \begin{pmatrix} 1 \\ 1 \\ 1 \end{pmatrix} = 1.$$

If the first layer is already filled , then for the second $(n = 2)$, for example, for $p = 0$, $q = 0$:

$$\begin{pmatrix} 2 \\ 0 \\ 0 \end{pmatrix} = \begin{pmatrix} 1 \\ -1 \\ 0 \end{pmatrix} + \begin{pmatrix} 1 \\ 0 \\ 0 \end{pmatrix} = 1, \text{ because from the conditions (8)} \begin{pmatrix} 1 \\ -1 \\ 0 \end{pmatrix} = 0, \text{ and } \begin{pmatrix} 1 \\ 0 \\ 0 \end{pmatrix} = 1,$$

and so on.

With each new passage of light  in the ray system in Fig. 1 new rays appear: inclined at greater angles, rays parallel to each other and rays superimposed on each other.

Rays having the same directivity, i.e. rays parallel to each other and overlapping  each other are inclined to the axis (Vertical in Fig. 1) at the same angle. These angles of ray inclination to the axis will be multiple of $\pm 2\gamma$ in each passage $n$.

The angular distribution of rays (Fig. 1) during  the passage $n$, i. e. the number of rays at angles multiple of $\pm 2\gamma$ or, to be more exact, $2(n - 2p)\gamma$ is equal to the sum of  numbers of row $p$ of layer $n$ (Fig. 4):

$$K_p = \sum_{q=0}^{n(n+1)/2} \begin{pmatrix} n \\ p \\ q \end{pmatrix} = \begin{pmatrix} n \\ p \end{pmatrix}, \tag{10}$$

i.e. is expressed by ordinary binomial coefficients. In Fig. 4 values $\begin{pmatrix} n \\ p \end{pmatrix}$ are the sums of rows $p$ in rectangles, are shown on the right of the rectangles in the first column. The second column contains the values of ray inclination angles for each passage $n$ (Fig.1), multiple of angles $\pm 2\gamma$, or, to be more exact, $2(n - 2p)\gamma$.

As mentioned earlier in Section 2, rays $K$ propagate along links $N$. The angular distribution of links (Fig. 1), i. e. the number of links at angles $2(n-2p)\gamma$ is

$$N_p = -p^2 + pn + 1. \qquad (11)$$

The quantity of links (Fig. 1) is the amount of natural numbers in $p$-row (Fig. 4).

Fig. 1 shows the points of branching $q$. The number of rays entering the points at different angles is shown in Fig. 4 as numbers in the vertical columns in triangles.

The number of rays going out of these branching points (Fig. 1) at different angles is twice as many than those entering is shown in triangles in Fig. 4 as columns of numbers inclined at 45°, the numbers are shown in bold type in triangles.

### 5. Non-linear arithmetic triangles

In Fig. 4 for $n=3$ and $n=4$ the sums of numbers in vertical columns and those inclined at 45° are shown in bold type, respectively; the sums are shown in lines below the rectangles; they show the overall amount of rays both entering and leaving the points of branching.

Let us write down separately all sums of numbers situated in vertical columns for all rectangles shown in Fig. 4. and distribute the rows of the sums in lines (analogously to Fig. 2b). As a result we will obtain a non-linear triangle of the first type (Fig. 5).

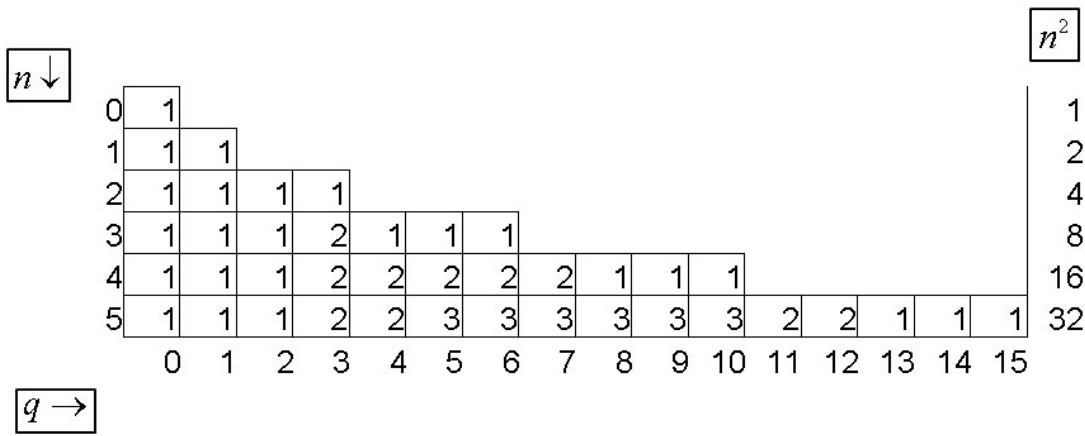

Fig. 5. The non-linear arithmetic triangle of the first type, the sum of numbers in rows is given in the column on the right. These numbers correspond to the number of rays entering the branching points $q$ (Fig. 1) at different angles during passing $n$.

Let us denote the number situated in $n$-row and $q$-column of the non-linear triangle as:

$$\begin{bmatrix} n \\ q \end{bmatrix} \qquad (12)$$

Let us set the initial conditions for the number of zero row $(n = 0)$:

$$\begin{bmatrix} 0 \\ q \end{bmatrix} = 1, \qquad (13)$$

for $q = 0$, i.e.

$$\begin{bmatrix} 0 \\ 0 \end{bmatrix} = 1 \text{ and } \begin{bmatrix} 0 \\ q \end{bmatrix} = 0 \qquad (14)$$

for the other $p$.

Let us also set the boundary conditions for the numbers $q$ of the other $n$-rows:

$$\begin{bmatrix} n \\ q \end{bmatrix} = 0 \qquad (15)$$

for $q < 0$, $q > n(n+1)/2$.

Then the rule of successive filling of the non-linear triangle with numbers beginning with the apex will be:

$$\begin{bmatrix} n \\ q \end{bmatrix} = \begin{bmatrix} n-1 \\ q \end{bmatrix} + \begin{bmatrix} n-1 \\ q-n \end{bmatrix}. \qquad (16)$$

**Example 1**. The number $\begin{bmatrix} 1 \\ 0 \end{bmatrix}$, i.e. for $n = 1$, $q = 0$, taking into account (13), (14), (15) will be equal to: $\begin{bmatrix} 1 \\ 0 \end{bmatrix} = \begin{bmatrix} 0 \\ 0 \end{bmatrix} + \begin{bmatrix} 0 \\ -1 \end{bmatrix} = 1$.

**Example 2**. The number $\begin{bmatrix} 1 \\ 1 \end{bmatrix}$, i.e. for $n = 1$, $q = 1$ will be:

$\begin{bmatrix} 1 \\ 1 \end{bmatrix} = \begin{bmatrix} 0 \\ 1 \end{bmatrix} + \begin{bmatrix} 0 \\ 0 \end{bmatrix} = 1$, and so on.

Now let us write down separately all the sums of numbers situated in the inclined columns shown in Fig. 4 and place the rows of the sums in lines. As a result, we will get a non-linear triangle of the second type (Fig. 6).

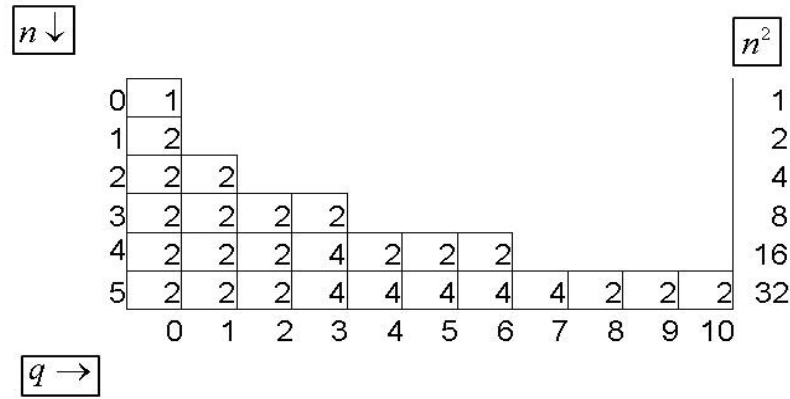

Fig. 6. The non-linear arithmetic triangle of the second type, the sum of numbers in the rows in the column on the right. These numbers correspond to the number of rays coming out of the points (Fig. 1) of branching $q$ at different angle during $n$-passage.

Let us denote the number located in $n$-row and $q$-column of the non-linear triangle of the second type as

$$\left\{ \begin{matrix} n \\ q \end{matrix} \right\} \qquad (17)$$

Let us set the initial conditions for the number of the zero row $(n = 0)$:

$$\left\{ \begin{matrix} 0 \\ q \end{matrix} \right\} = 1, \qquad (18)$$

for $q = 0$, i.e.

$$\left\{ \begin{matrix} 0 \\ 0 \end{matrix} \right\} = 1 \text{ and } \left\{ \begin{matrix} 0 \\ q \end{matrix} \right\} = 0 \qquad (19)$$

for the other $p$.

Let us set also the boundary conditions for numbers $q$ of the other $n$-rows:

$$\left\{ \begin{matrix} n \\ q \end{matrix} \right\} = 0 \qquad (20)$$

for $q < 0, \ q > n(n-1)/2$.

Then the rule of successive filling with numbers of the non-linear triangle beginning with the apex will be:

$$\left\{ \begin{matrix} n+1 \\ q \end{matrix} \right\} = \left\{ \begin{matrix} n \\ q \end{matrix} \right\} + \left\{ \begin{matrix} n \\ q-n \end{matrix} \right\}. \qquad (21)$$

**Example 3**. The number $\left\{\begin{matrix} 1 \\ 0 \end{matrix}\right\}$ for $n+1=1$, $q=0$, taking into account (18), (19), (20), will be equal to: $\left\{\begin{matrix} 1 \\ 0 \end{matrix}\right\} = \left\{\begin{matrix} 0 \\ 0 \end{matrix}\right\} + \left\{\begin{matrix} 0 \\ 0 \end{matrix}\right\} = 2$.

**Example 4**. The number $\left\{\begin{matrix} 1 \\ 1 \end{matrix}\right\}$, for $n+1=1$, $q=1$ will be equal to: $\left\{\begin{matrix} 1 \\ 1 \end{matrix}\right\} = \left\{\begin{matrix} 0 \\ 1 \end{matrix}\right\} + \left\{\begin{matrix} 0 \\ -1 \end{matrix}\right\} = 0$, and so on.

In this case expressions (16) and (21) for non-linear triangles are not equivalent to each other unlike (4) for the linear case.

Notations (12) and (17) differ from the ordinary notation of the binomial coefficient, i.e. for our case:

$$\begin{bmatrix} n \\ q \end{bmatrix} \neq \begin{pmatrix} n \\ q \end{pmatrix} \text{ and } \left\{\begin{matrix} n \\ q \end{matrix}\right\} \neq \begin{pmatrix} n \\ q \end{pmatrix}, \text{ where } \begin{pmatrix} n \\ q \end{pmatrix} = \frac{n!}{q!(n-q)!} \text{ [7]}.$$

Expression (16) corresponds to the known formula [8] of integers partition. The number (12) shows by how many method the number $q$ may be partitioned into items each of which is equal to one of the numbers $1,2,3,\ldots,n$, the order of items being not taken into account and all items being different. Thus, for example, number 7 may be partitioned into items $1,2,3,4,5$ by three methods: $7 = 3+4$, $7 = 2+5$, $7 = 1+2+4$; i.e. for our case for $n=5$, $q=7$: $\begin{bmatrix} 5 \\ 7 \end{bmatrix} = 3$ (Fig. 5) and so on.

Expressions(4), (9) and (21) agree with the formulas for other types of partitions.

In Fig.3 $p$, the pyramid side, is the numbers composing Pascal triangle and $q$, the side, is the numbers composing a non-linear triangle of the first type. Analogously, a non-linear arithmetic pyramid may be constructed where $p$, the pyramid side, is the numbers composing Pascal triangle and $q$ is the numbers composing a non-linear triangle of the second type.

### 6. Common properties of the ray system and the arithmetic pyramid

The ray system (Fig. 1) and, correspondingly, the non-linear arithmetic pyramid (Fig. 3, 4) have the following common features:

1. The amount of rays $K$ in each passage $n$ (Fig. 1) and the sum of numbers in the layer $n$ (Fig. 4) is equal to $2^n$, coincides with the sum of numbers in row $n$ of Pascal triangle (Fig. 2) and with the sum of numbers in row $n$ of non-linear triangles.

2. The amount of rays $K$, inclined at the same angles in each passage of $n$ (Fig. 1) and the sum of numbers in line $p$ of layer $n$ (Fig. 4) is equal to $\binom{n}{p}$ – corresponds to binomial coefficient (member $p$ of row $n$ (Fig. 2) of Pascal triangle).

3. The amount of different angles multiple of $2\gamma$ at which rays in each passage $n$ are inclined (Fig. 1) and the height $p+1$ of the triangle of layer $n$ (Fig. 3, 4) is equal to $n+1$, i.e. is expressed by natural row: $1,2,3,4,5,\ldots,n+1$ and coincides with the amount of numbers in row $n$ (Fig. 2) of Pascal triangle.

4. The amount of points $q$ of ray branching after each passage $n$ (Fig. 1) and length $q+1$ of the layer $n$ rectangle $n$ (Fig. 4) is expressed by the row: $1,2,4,7,11,16,\ldots,\ n(n+1)/2+1$, coincides with the amount of numbers in row $n$ of the non-linear triangle of the first type (Fig5).

5. The amount of links $N$ along which $K$ rays are distributed in each $n$-passage (Fig. 1) and the amount of natural numbers within the limits of rectangle layer $n$ (Fig. 4) is expressed by the row: $1,2,4,8,15,26,\ldots,$ $n(n^2+5)/6+1$.

A special feature of the above last three rows is the fact that the difference of the numbers of the row given in item 5 is the row given in item 4 and the difference of the numbers of the row given in item 4 is the row given in item 3:

$$1,\ \ 2,\ \ 4,\ \ 8,\ \ 15,\ \ 26,\ \ 42,\ \ldots$$
$$1,\ \ 2,\ \ 4,\ \ 7,\ \ 11,\ \ 16,\ \ldots \tag{22}$$
$$1,\ \ 2,\ \ 3,\ \ 4,\ \ \ 5,\ \ldots$$

It is possible to present new parallels between the properties of the ray system and those of the non-linear arithmetic pyramid.

Our numerical calculations based on the study of the arithmetic pyramid show that when number $n$ increases the envelopes of angular distribution of rays, entering and leaving each of the branching points as well as the angular distribution of rays of the whole system (Fig. 1) tend to normal.

Transverse distribution (within the laser aperture) of the total amount of rays in the points of branching also tends to normal.

### 7. Conclusion

It is seen from the above illustrative geometrical models that for the case of small angles the angular distribution of the amount of rays after $n$ passages grows linearly as $p$-pyramid sides (Fig. 3), while the diameter of the beam itself grows considerably faster, non-linearly, as $q$-sides of the pyramid, therefore the beam, expanding, rapidly leaves the limits of the laser aperture.

The consideration of ray distribution at the points of their branching makes it possible to demonstrate not only the angular distribution of light when it leaves the laser but also the distribution of the light field in the laser itself.

The geometric approach may be used for the study of the properties of the beams of broad-aperture and multi-frequency lasers, as well as for the description of the distribution of laser pulses among the mirrors of the laser resonator.

### Acknowlegement